\documentclass{article}
\usepackage{authblk}
\usepackage{graphicx}
\usepackage{cite}

\begin{document}

\title{Trend and Fractality Assessment of Mexico's Stock Exchange}

\author[1]{Javier Morales \thanks{tequilaydiamante@uanl.edu.mx}}

\author[2]{V\'ictor Tercero 
\thanks{victor.tercero@itesm.mx}}

\author[1]{Fernando Camacho \thanks{jose.camachovl@uanl.edu.mx}}

\author[1]{Eduardo Cordero 
\thanks{lalo.cordero@gmail.com}}

\author[1]{Luis L\'opez 
\thanks{nerioluisnerio@gmail.com}}

\author[1]{F-Javier Almaguer 
\thanks{FRANCISCO.ALMAGUERMRT@uanl.edu.mx}}

\affil[1]{Universidad Aut\'onoma de Nuevo Le\'on}
\affil[2]{Tecnol\'ogico de Monterrey}

\date{\today}
\maketitle

\begin{abstract}
The total value of domestic market capitalization of the Mexican Stock 
Exchange was calculated at
520 billion of dollars by the end of November 2013. To manage this system 
and make optimum
capital investments, its dynamics needs to be predicted. However, 
randomness within the stock indexes
makes forecasting a difficult task. To address this issue, in this work, trends and 
fractality were studied using $GNU-R$ over the opening and closing prices indexes over the past 23 years. 
Returns, Kernel density estimation, autocorrelation function and $R/S$ analysis
and the Hurst exponent were used
in this research. As a result, it was found that 
the Kernel estimation density and the autocorrelation function shown the presence of long-range memory effects. In a first approximation, the returns of closing prices seems to behave
according to a Markovian random walk with a length of step size  given by an alpha-stable random process. For extreme values, returns decay asymptotically as a power law with a characteristic exponent approximately equal to 2.5.
\end{abstract}


\section{Introduction}
\label{1}

Financial markets exibit a dynamic behaviour in the form of fluctuations, trends, and volatility.
Market regulations, globalization, changes in the interest rates, war conflicts, new technologies,
social movements, news and housing are only a small sample of factors affecting the chaotic and complex structure of the financial markets \cite{cruz, borrius, kose, aguiar, neumeyer, guidolin, schneider, dicle, king, boyd, engle, piskorec, case}. 

To assess all these interacting elements within a coherent theoretical economic framework to create
a prediction model is, at least for now, a nearly impossible task. The behaviour of 
an economic system integrates a collection of emerging properties of chaotic and complex systems \cite{holyst, hsieh, may, peinke, bonanno, stauffer, sornette1, hens}.
From a deteministic approach, the effort requiered to model and characterize a such system might be
monumental. Complex correlations in the fluctuations of financial and economic indices \cite{preis1, wang, preis2, pasquini}, the self-organization
phenomena in market crashes \cite{focardi, turcotte, huang, delbrio}, the sudden high growth or sharp fall in the stock market during periods of apparent stability \cite{veldkamp, boldrin, sornette2, johansen}, are only a few examples
of situations that can be considered difficult to understand or represent mathematically.

On one hand, from the stochastic-deterministic point of view of the physical statistics \cite{voit1, plerou, mantegna1, mantegna2, ingber, voit2},
the dynamic variation of prices in a financial market can be considered as a result of an enormous
amount of interacting elements. For instance, stock prices are the result of multiple increments
and decrements that result from a feedback response of every action defining the composition of an index, which result, at the same time, from decisions and flows of information that change
from one moment to the next \cite{kwon, mitchell, zhang, marschinski, french}. For this reason, a detailed description of each trajectory, 
within the structure of a system, would be almost imposible and futile: the series of events that gave birth
to a specific trajectory might not repeat, and a detailed description would not have a
predictive utility.

On the other hand, by considering axiomatic that we are unable to reach a deterministic understanding
of a system as whole, a statistical approach might be useful tu describe the uncertainty involved.
At least we would be able to gain some insight about the expected behaviour, size of fluctuations, or
the corresponding probabilities of rare events. This, with the final intention of doing forecasts
on the process future behaviour. The statistical description may even predict in an essentially deterministic way, such as the diffusion equation which describe the density of particles, each one performing a random walk in a microscopic scale \cite{bassler, sokolov, metzler}.

According to web portal of Mexican Stock Exchange, ``The Prices and Quotations Index (MEXBOL) 
is the Mexican Market’s main indicator, it expresses the stock market return according to 
the prices variations of a balanced, weighted and representative Constituent List of 
the equities listed in the Mexican Stock Exchange, in accordance to best practices
internationally applied'' \cite{bmv}. The daily values of the MEXBOL or IPC Index form a set of variables: The Opening Price, the Closing Price, the High an Low values, the Adjusted Price and the transaction Volume. The Fig. 1 show the variation in time of the closing price $\{Y_{n}, n = 0, 1, 2, \cdots, N\}$, $N = 5719$.

\begin{center}
\begin{figure}[htbp] 
\begin{center}
\includegraphics[scale = 0.30, angle = 0]{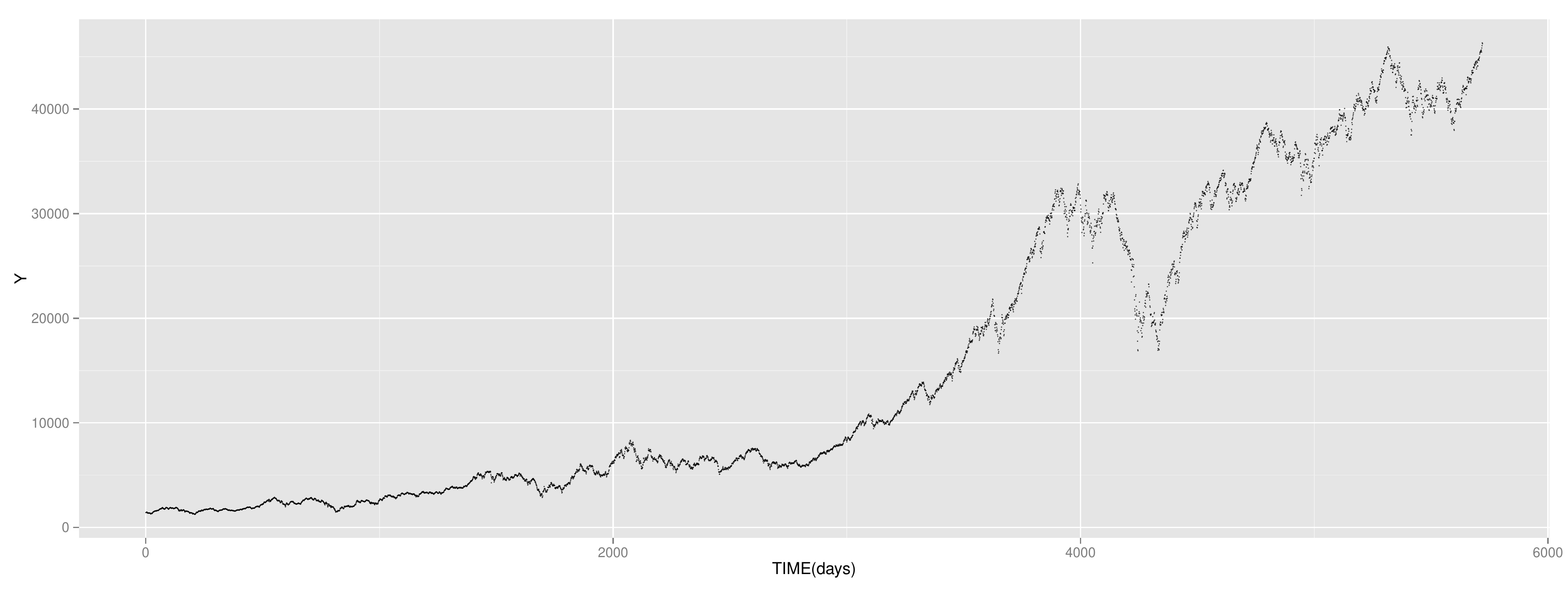}
\end{center}
\caption{MEXBOL. Closing price of the IPC Index from 1991, November 8 to 2014, September 5. Coarse-grained approach: Despite of presence of at least three 
strong financial crisis (see Fig. 10 in section conclusions), the closing price, $\{Y_{n}, n = 0, 1, 2, \cdots, N\}$, $N = 5719$, show an ascendant 
``convex'' trend until about July of 2007; 
from March of 2009 initiates a ``concave'' sustained growth. 
The economic ``interregnum'' between July/2007 and 
March/2009 marks the transition between the two regimes.\label{IPC}}
\end{figure}
\end{center}

\section{Methodology}
\label{2}

To perform this research, 23 years of observations from MEXBOL were used, from November the 8th, 1991, to September the 5th, 2014.
This information is publicly available in several websites, see for instance the historical prices of IPC(\textasciicircum MXX) or IPC Index-Mexico in \cite{ipc}.

This paper presents an assessment of the dynamic behaviour of closing and opening princes of the IPC Index from three different perspectives:
a) Analysis of the stochastic properties of the random behaviour and fluctuations of returns of closing price.
b) Estimation of the degree of fractality and long term memory through the rescaled range or $R/S$ analysis and the Hurst exponent.
c) Empirical autocorrelation function analysis. 

All the analysis and the numeric and visual calculations of the IPC Index properties
were done using the GNU-R free software environment within an Ubuntu-Linux 14.04 work enviroment.

\subsection{Closing prices}
The return values is a regular transformation used in economic data in order to standardize and remove the trending. 
Thus, a simple and fast way to detrending the closing price $\{Y_{n}, n = 0, 1, 2, \cdots, N\}$ serie, Fig. \ref{IPC}, 
is given by the transformation

\begin{equation}\label{return}
r_{n} = \frac{Y_{n}}{Y_{n-1}}, \;\;\; n = 1, 2, \cdots, N.
\end{equation}

The mean value of this new series is $\langle r\rangle = 1.000729$, i.e., the  daily closing price has a rate of return average of $\langle \Delta Y/Y\rangle = \langle r\rangle - 1 = $ 7.29 per 1000 units.
The corresponding standard deviation is $\langle\langle r\rangle\rangle = 0.01548542$. 
A comparison between normalized logarithms of returns (mean zero and variance 1) with the profile created by a Gaussian white noise, also with mean zero and variance 1, is shown in Fig. 2.

\begin{center}
\begin{figure}[htbp] 
\begin{center}
\includegraphics[width = 4.9in]{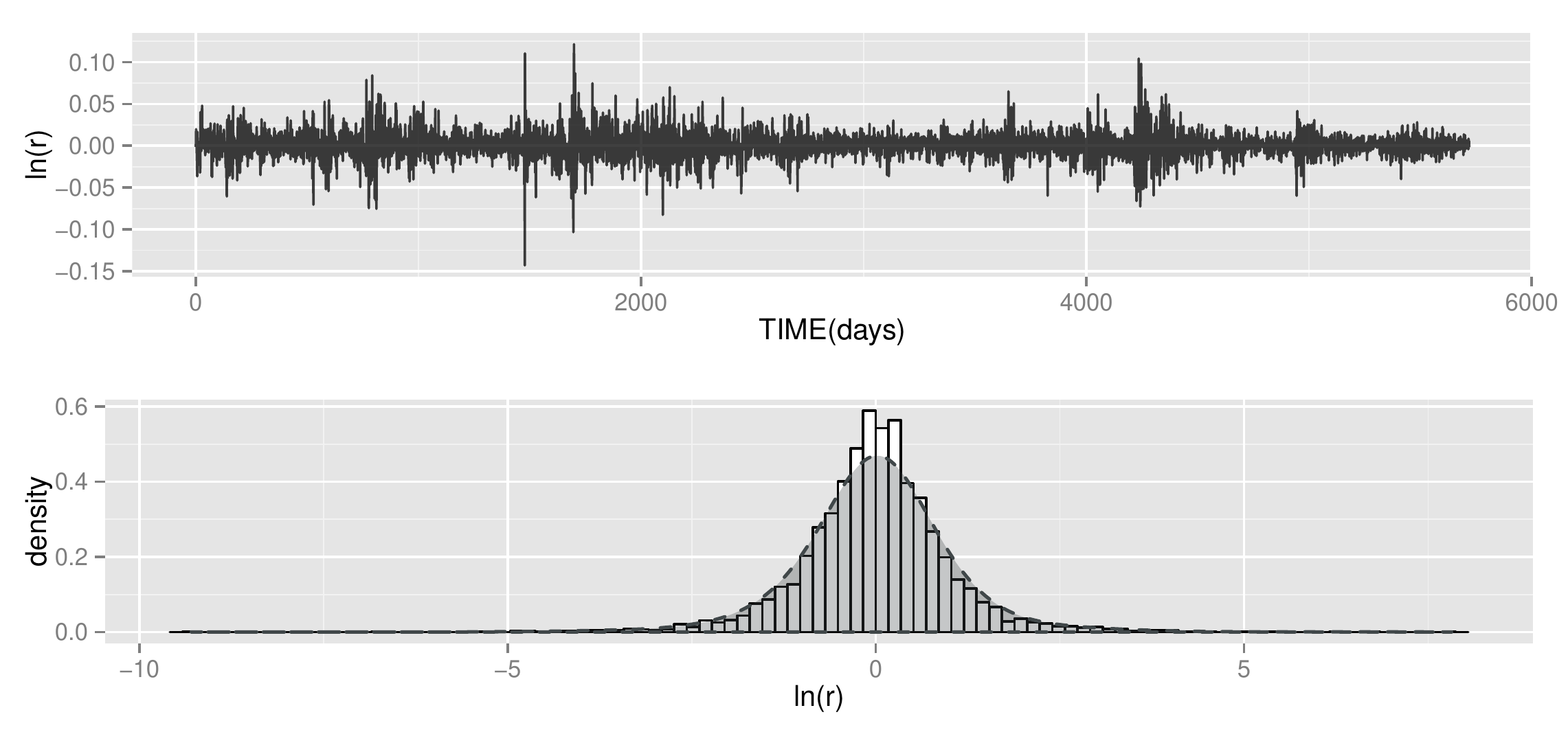}
\end{center}
\caption{Top: Normalized logarithmic returns of the closing price $\{Y_{n}\}$ serie for the IPC Index of Fig. 1. Bottom: The corresponding histogram 
for the normalized logarithmic returns
of the closing price for the IPC Index. The quantiles corresponding to the area under 
the curve  $(0, 25\%, 50\%, 75\%, 100\%)$ are $(-9.30, -0.49, 0.01, 0.50, 7.82)$, approximately. \label{returns}}
\end{figure}
\end{center}

\begin{center}
\begin{figure}[htbp] 
\begin{center}
\includegraphics[width = 4.9in]{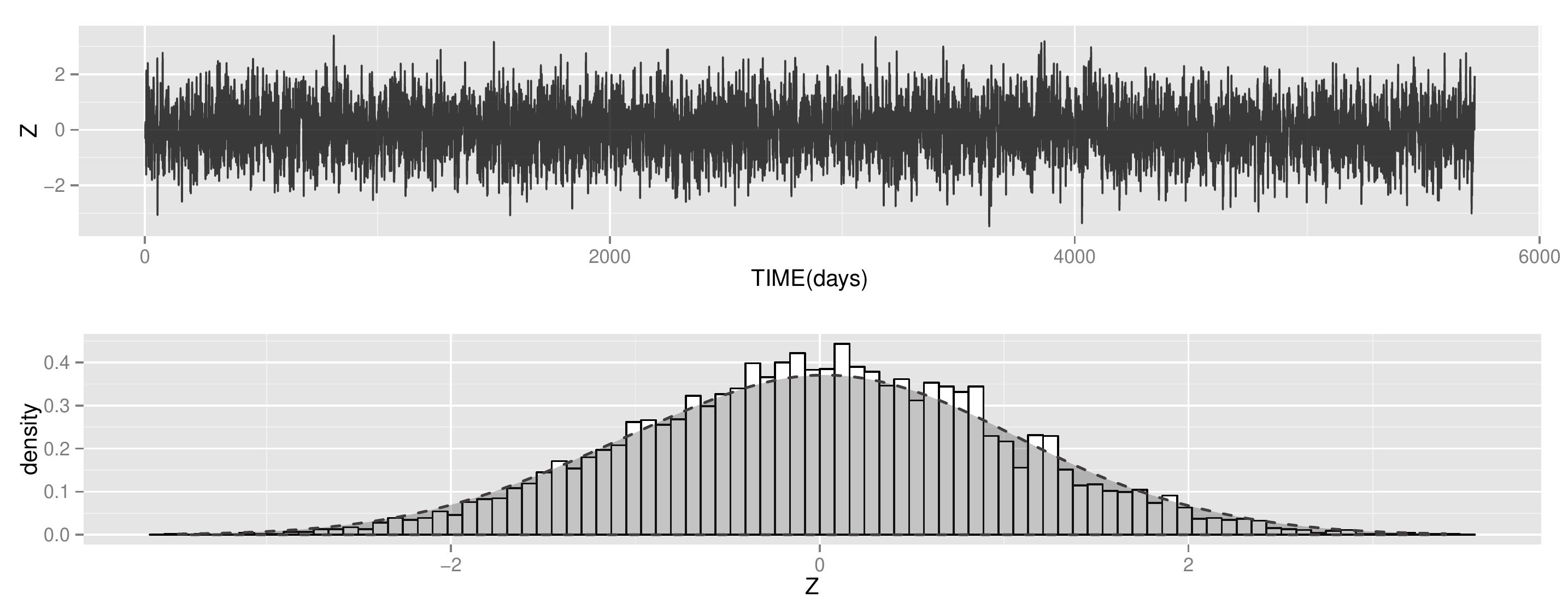}
\end{center}
\caption{Top: Empirical time serie for normalized Gaussian white noise. Bottom: The corresponding histogram for the normalized Gaussian white noise
values with typical quantiles $(-3.83, -0.66, -0.01, 0.66, 3.47)$ for $(0, 25\%, 50\%, 75\%, 100\%)$ area under the curve. \label{gauss}}
\end{figure}
\end{center}

\noindent According to Figs. \ref{returns} and \ref{gauss}, unlike the normalized Gaussian white noise, the distribution of returns or logarithmic returns is more narrower with
a higher concentration of values in $\pm 1\sigma$ from the mean but with a tail somewhat heavier. 

From a sample or time serie $\{X_{i}, i = 1, 2, \cdots, N\}$, an empirical probability density distribution ${\hat P}(x)$ is given by the superposition of normalizated kernels through the Kernel Density Estimator or KDE approximation \cite{sheather}

\begin{equation}
{\hat P}(x) = \frac{1}{Nh}\sum_{i = 1}^{N}K\left(\frac{x-X_{i}}{h}\right)
\end{equation}

\noindent where $h$ is the bandwidth or scale parameter and $K(\cdot)$ is a normalizated kernel function. 

Using the library \emph{kedd} in \emph{GNU-R} \cite{guidoum}, an empirical probability density distributions KDE-based with an optimal bandwidth for both, normalized Gaussian white noise and normalized logarithmic returns, are obtained. In the Fig. 4 the sharpe fall of $log(r)$ respect the normalized
Gaussian white noise becomes evident.

\begin{center}
\begin{figure}[htbp] 
\begin{center}$
\begin{array}{cc}
\includegraphics[width = 4.7in]{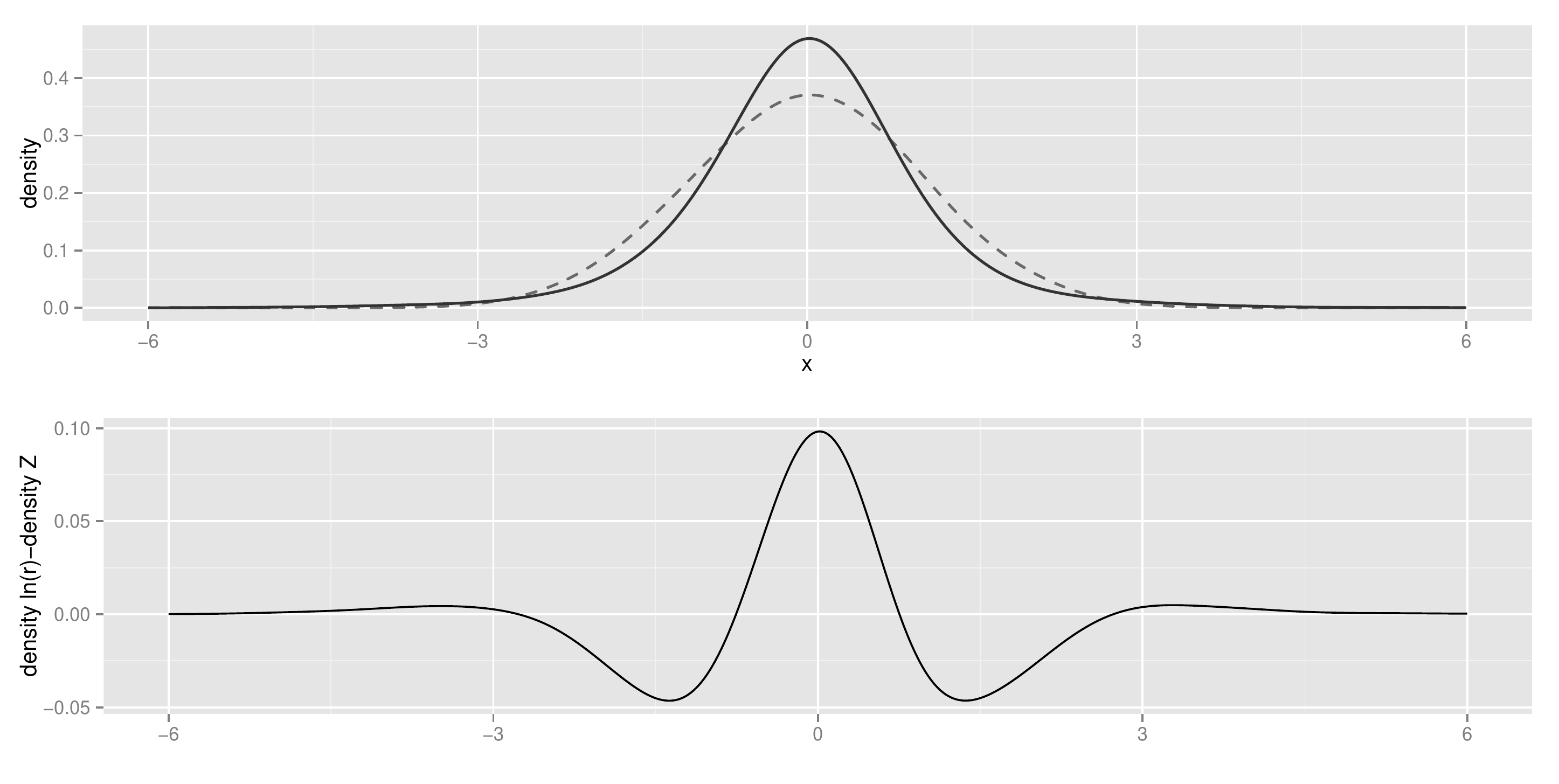}&
\end{array}$
\end{center}
\caption{Top: Kernels estimation density $K_{r}(x)$ (continuous line) and $K_{z}(x)$ (dashed line) for normalized, logarithmic returns and Gaussian white noise, respectively, with bandwidth $h = 0.4044283$, and an Asymptotic Mean Integrated Squared Error (AMISE) of $\approx 0.0003$. Bottom: Difference $\Delta K(x) = K_{r}(x)-K_{z}(x)$ between both densities. Observe that, roughly when $x \in [-2.71, -0.71]\cup [0.81, 2.77], \Delta K(x) < 0$ and the area below the Gaussian curve is greatest that the area below the curve of $log(r)$ values. However, in the central region $x \in [-0.71, 0.81], \Delta K(x) > 0$ and the $log(r)$ values are more concentrated respect the Gaussian values. Finally, in the tails $x \in [-6.0, -2.71]\cup[2.77, 6.0]$ the distribution of the $log(r)$ is more heavy that the Gaussian white noise.\label{kernel-density}}
\end{figure}
\end{center}

From the Fig. \ref{kernel-density}, for values approximately
in $0.75\sigma$ around the mean, the distribution of $log(r)$ is more sharply. On the other hand, approximately when
$\mid x\mid > 2.7 \sigma$ heavy tails emerge and this indicates the presence of a coarse-grained power law underlying the dynamics of return fluctuations, very similar to the fluctuations in random variables with an $\alpha-stable$ L\'evy distributions $S_{\alpha}(x; \beta, \gamma, \delta)$ characterized by $\alpha$, 
$\beta$, $\gamma$ and $\delta$, the stability, skewness, scale, and location parameters, respectively \cite{samorodnitsky}. When $x\gg 0$, the density probability $S_{\alpha}$ decays asymptotically as \cite{wertz}

\begin{equation}\label{stable}
f(x) \sim \alpha(1+\beta)C_\alpha x^{-(1 + \alpha)}
\end{equation}

\noindent with $C_{\alpha} = \Gamma(\alpha)[\pi \sin(\alpha\pi/2)]^{-1}$.

\emph{A comment}: Assuming that $f_{r}(x)$ is the density probability function of the closing prices $\{Y_{n}\}$, the transformation $\{r_{n}\} \rightarrow \{R_{n}=\ln(r_{n}) \}$, $n = 1, 2, \cdots, N$, 
define a density probability for the logarithmic returns $f_{R}(x)$ showing an asymptotic fall with a slightly higher heavy-tail such that $f_{R}(x) = f_{r}(r(R))e^{R} > f_{r}(x)$. 

\subsection{Opening and Closing prices}
To concentrate on analyzing the \emph{financial noise}, another way to detrending a time series is to obtain the first differences or \emph{jumps} to \emph{nearest neighbors}. So some examples of this transformation are: the first differences within the closing prices serie $\{W_{n} = Y_{n}-Y_{n-1}\}$, or the first differences delayed one day between the opening and closing price series $\{L_{n} = X_{n}-Y_{n-1}\}$, both with $n = 1, 2, \cdots, N$. 

A graph with the differences $\{L_{n}\}$ between the opening prices of $(n)$-th day and the $(n-1)$-th closing price of the day
before can be observed in Fig. 5 in top. Here, much more than the logarithmic returns series the appearance of \emph{extreme events} is more evident. The mean and standar deviation values for this serie are $\langle L\rangle \approx 0.70$ and $\langle\langle  L\rangle\rangle \approx 19.29$, respectively. 

A special case of a transformation by differences are the first differences day by day between the closing and opening prices serie $\{D_{n} = Y_{n}-X_{n}, \; n = 0, 1, 2, \cdots, N\}$. Particularly, this serie shows a symmetrical pattern around the mean value
$\langle D\rangle \approx 7.14$ with a standard deviation of $\langle\langle D\rangle\rangle \approx 242$. Also, although without apparent trend shows a high volatility increasing in time, see Fig. 5 in bottom. 

\begin{center}
\begin{figure}[htbp] 
\begin{center}$
\begin{array}{cc}
\includegraphics[width = 4.7in]{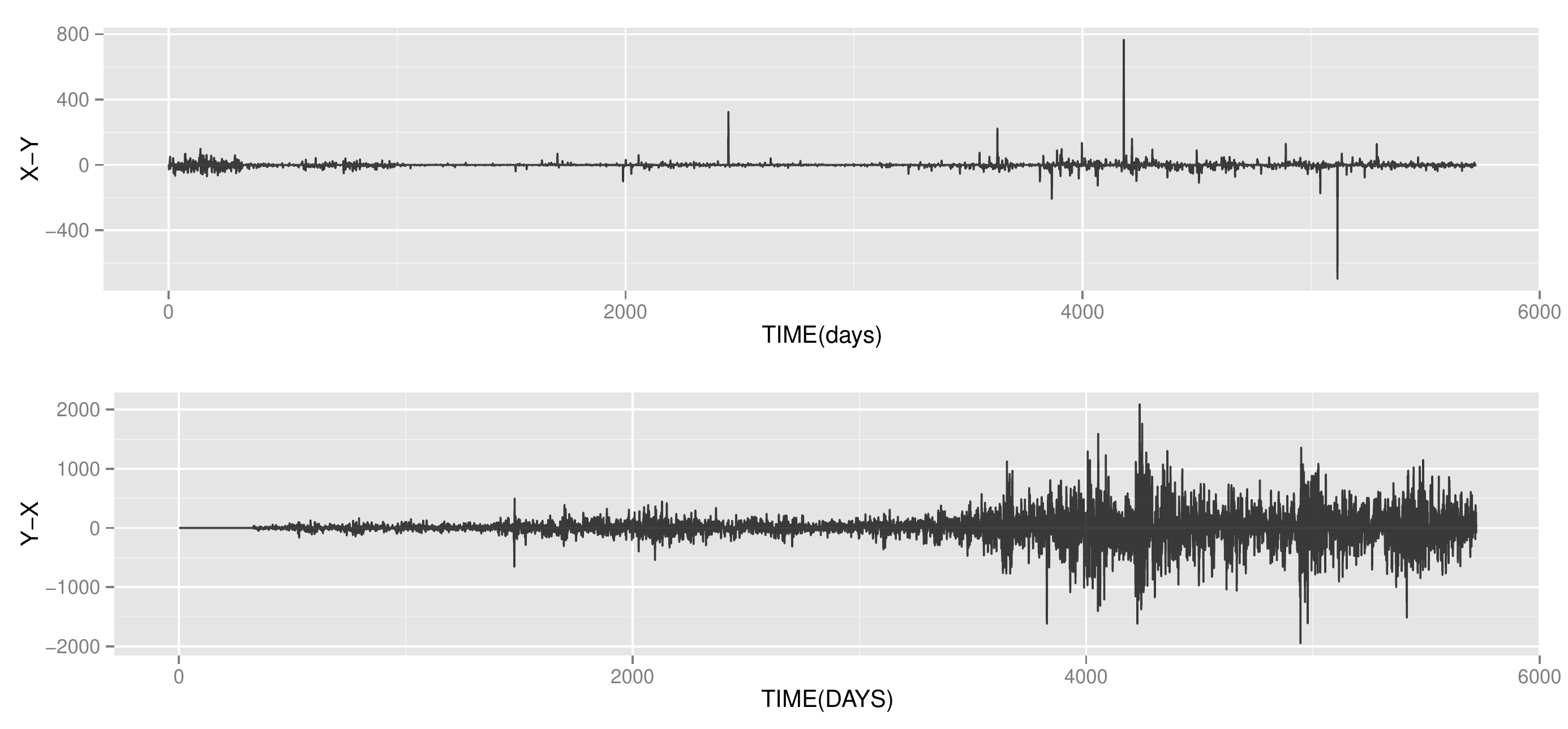}&
\end{array}$
\end{center}
\caption{Top: The normalized differences  $\{L_{n} = X_{n}-Y_{n-1}, \;n = 1, 2, \cdots, N\}$ between the opening price in the $(n)$-th day with the closing price in the $(n-1)$-th day. The quantiles $(0\%, 25\%, 50\%, 75\%, 100\%)$ for this serie are $(-39.7, -0.06, -0.04, 0.06, 36)$ approximately. Bottom: The normalized differences $\{D_{n} = Y_{n} - X_{n}, \; n = 0, 1, 2, \cdots, N\}$ between the closing and opening prices for the IPC Index in the $n$-th day. The quantiles $(0\%, 25\%, 50\%, 75\%, 100\%)$ for the normalized differences $\{D_{n}\}$ are $(-8.08, -0.26, -0.029, 0.29, 8.60)$ approximately. \label{difference}}
\end{figure}
\end{center}

Unlike a Gaussian white noise, one of the main features of $\alpha-$stable distribution is the existence of long-term correlations. A measure of the intensity of this \emph{memory effect} and the fractal behavior type of fluctuations in a stochastic
process is given by the value of the Hurst exponent. The Hurst exponent can be obtained by the rescaled range analysis, or $R/S$ analysis,
a standard technique to measure the fractality or the persistence in time of correlations in time series. In the next section, this technique is applied to the different series presented so far.

\subsection{Hurst exponent}

The Hurst exponent, a parameter introduced in 1951 by the British hydrologist 
Harold Edwin Hurst to study the Nile river risings, is an instrument used to 
measure at different scales the mean intensity of the fluctuations of a time series and its 
tendency to form \emph{clusters of persistence}. 
For construction, the Hurst 
exponent (henceforth $H$ parameter) is a real number between $0$ and $1$. 
If $0.5 < H \leq 1$ it is said that the mean tendency of the series is 
persistent (either persistently upwards or downwards). If $0 \leq H < 0.5$ 
it is said that the series is anti-persistent 
(it goes from upwards to downwards or vice-versa). Finally, 
when $H = 0.5$ we can say that in average there is no persistence, 
and then it is a memory-less process: the values of the process are completely 
uncorrelated and the behaviour of the series is the one observed in 
a Gaussian white noise random process or like the uncorrelated incremets in the  Brownian motion process \cite{carbone}. 

The strong trend observed in the closing prices of Fig. 1 is the result of the persistence of high levels of correlation in time.
In principle, the Hurst exponent $H$ allows to differentiate between series of Gaussian white noise created by a series of independent random variables as the increments of 
an ordinary Brownian motion, i.e., between a stochastic processes with a future completely determined by the present (current state) and a more complex movement where the
\emph{quality} associated with the estimation of future values depends, in principle, of all previous observations. 

\subsection{Rescaled range analysis}
In the $R/S$ analysis the time series $\{X_{n}, n = 1, 2, \cdots, N\}$ is divided into $d$ sub-series, each of approximate size $m = N/d$. A heuristic form to take the partition of the serie is in power series of two, 
in this case the values of the index $d$ and $m$ are, respectively, 
$d = 1, 2, 4,\cdots, 2^K$, for some integer $K$,  and $m = N/1, N/2, N/4, \cdots, N/ 2^{K}$.
For each $d$ sub-series find the mean $E_{\ell}$ 
and the standard deviation $S_{\ell}$, $\ell = 1, 2, \cdots, d$.
Data are normalized by substracting the mean of each sub-serie. 
With the cumulative deviations 
$Y _{j} = \sum\limits_{i=1}^{m}(X_{i}-E_{l})$, $j = 1, 2, \cdots, m$; $l = 1, 2, \cdots, d$
the range $R_{\ell} = max(Y_{j})-min(Y_{j})$ is obtained and it is 
rescaled by the standar deviation $R_{\ell}/S_{\ell}$, $\ell = 1, 2, \cdots, d$. 
The mean of rescaled range for each partition, in the time scale $d$, is $\langle R/S\rangle_{d} = (1/d)\sum_{\ell = 1}^{d}$
$R_{\ell}/S_{\ell}$. Repeat the same with other 
partition for another $d$ value. Identifying $d \equiv t$, Hurst found that $\langle R/S\rangle_{t}$ scales by power-law as time 
increases, i. e., $\langle R/S\rangle_{t} = ct^{H}$
where $c$ is a constant. In practice $H$ can be estimated as 
the slope of a $log/log$ plot of $\langle R/S\rangle_{t}$ versus length $t$.

\subsection{Autocorrelation analysis}

It is said to a stochastic process is weakly stationary or wide-sense stationary if the function of mean is constant in time and the autocovariance
function $\gamma(t_{1}, t_{2})$ only depends on the elapsed time $\mid t_{2}-t_{1}\mid$ \cite{lefebvre}. 
Assuming the foregoing, the empirical autocovariance function for a finite sample of $N$ values of a wide-sense stationary time series
$\{X_{t}, t = 1, 2, \cdots, N\}$ can be estimated by \cite{shumway}

\begin{equation}\label{autocorrelation}
\hat{\gamma}(s, \tau)\equiv\hat{\gamma}(h) = \frac{1}{N}\sum_{t=1}^{N-h}(X_{t+h}-\overline{X})(X_{t}-\overline{X})
\end{equation}

\noindent with $h = \mid \tau-s\mid$ and $\overline{X}$ the common empirical mean. And the empirical autocorrelation function is

\begin{equation}\label{autocorrelation}
\hat{\rho}(h) = \frac{\hat{\gamma}(h)}{\hat{\gamma}(0)}, \;\;\;\;\;\;\;\; -1\leq \rho(h) \leq 1.
\end{equation}

\noindent with $\hat{\gamma}(0) = Var(X)$ the variance of the time serie $\{X_{t}, t = 1, 2, \cdots, N\}$. 

The autocorrelation function for the returns dies quickly after a lag of 1 into a pattern in appearance very similar to a Gaussian white noise, Fig. 8. 

\begin{center}
\begin{figure}[htbp]
\begin{center}$
\begin{array}{cc}
\includegraphics[width = 4.7in]{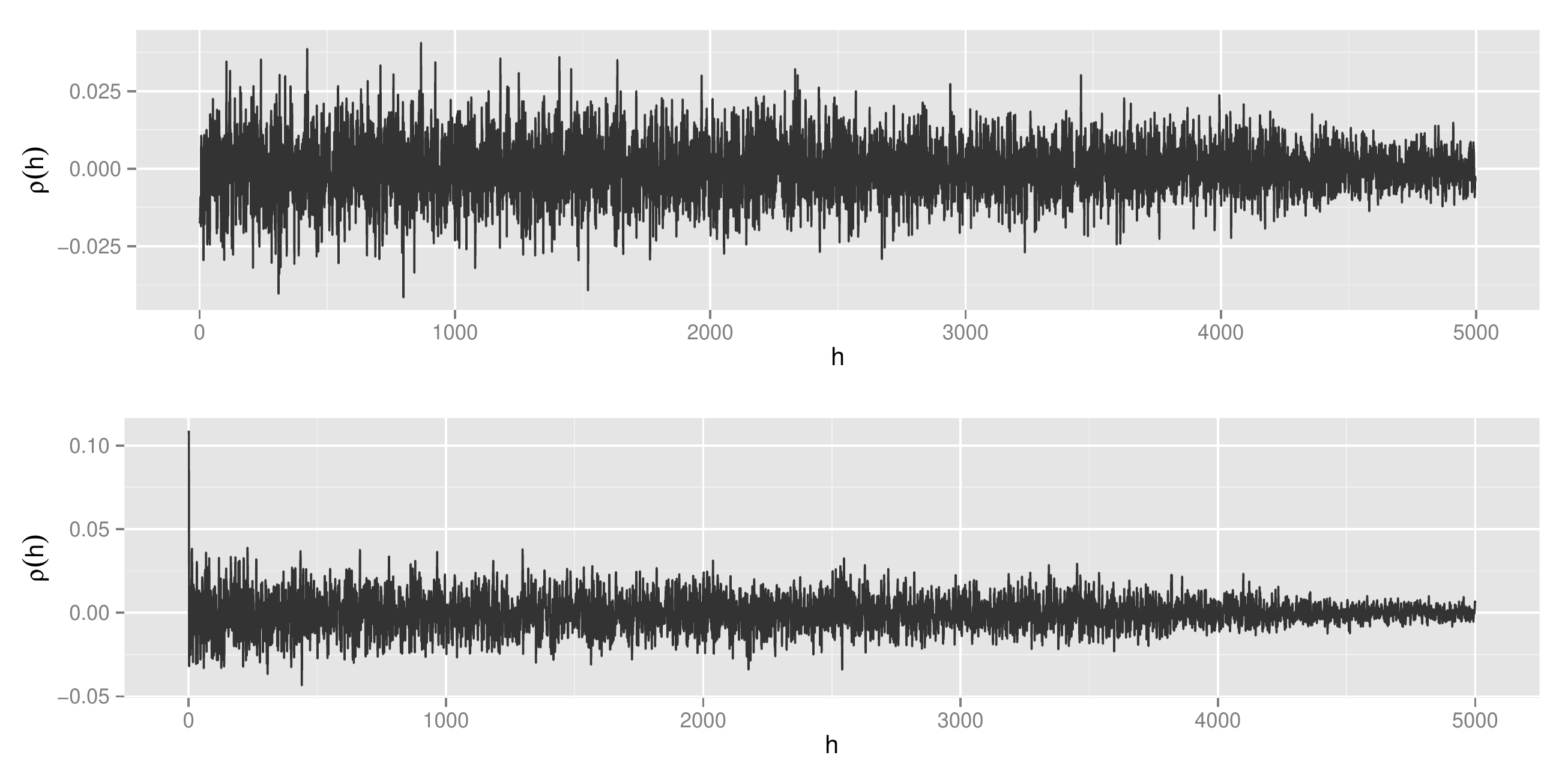}&
\end{array}$
\end{center}
\caption{The autocorrelation function (\ref{autocorrelation}) as a function of lag $h$. Top: The typical profile of $\rho(h)$ for a sample of Gaussian White noise; the mean and the standar deviation are $\langle \rho\rangle = -8.7\times 10^{-5}$ and $\langle\langle  \rho\rangle\rangle = 0.0093$, respectively. Bottom: The autocorrelation function for the  normalized returns of closing price $\{r_{n}\}$, mean value $\langle \rho\rangle = -8.9\times 10^{-5}$ and standar deviation $\langle\langle \rho\rangle\rangle = 0.0138$. The autocorrelation function presents a sharp fall for $h > 1$. \label{difference}}
\end{figure}
\end{center}

The progressive thinning of the autocorrelation function (an biased estimator) for a Gaussian white noise shown in Fig. 8 is a \emph{border effect} 
due to the scaling factor $1/n$ in (\ref{autocorrelation}). In fact for a sample Gaussian white noise the distribution of the autocorrelation function $\rho(h)$ for a maximum $h$ value is Gaussian approximately with mean $\langle \rho\rangle = 0$ and standar deviation decaying as $\langle\langle \rho \rangle \rangle = n^{-1/2}$ \cite{shumway}.

Nevertheless, unlike the Gaussian white 
noise, the fluctuations of the autocorrelation function for the returns displays a strong correlation to first neighbors. Moreover the maximum value in the autocorrelation function for the returns of any order occurs for a lag of $h = 1$.  
In other words, the transition of the returns $\{r_{n}\}$ to a symmetrical non Gaussian autocorrelation function for $h > 1$, Fig. 8 in bottom, 
suggest as a first approximation the following simple Markovian random walk for the returns of closing prices 

\begin{equation}
r_{t+1} = r_{t} + \xi_{t}
\end{equation}

\noindent where the noise $\xi_{t}$ is an $\alpha-$stable  random noise whose distribution has a stability parameter $\alpha < 2$ (non Gaussian), see Table 1.

Additionally the behavior of the autcorrelation function (\ref{autocorrelation}) for the first differences of $\{L_{n} = X_{n}-Y_{n-1}\}$ and $\{D_{n} = Y_{n}-X_{n}\}$ series were estimated. The landscape of the autocorrelation function for the first differences $\{L_{n}\}$ show correlations more complex at different scales, Fig. 7 on the left side. On the other hand, the profile of the autocorrelation function for the differences $\{D_{n}\}$ are  correlated beyond the first neighbors, Fig. 7 on the right side.

\begin{center}
\begin{figure}[htbp] 
\begin{center}$
\begin{array}{lr}
\includegraphics[width = 2.3in]{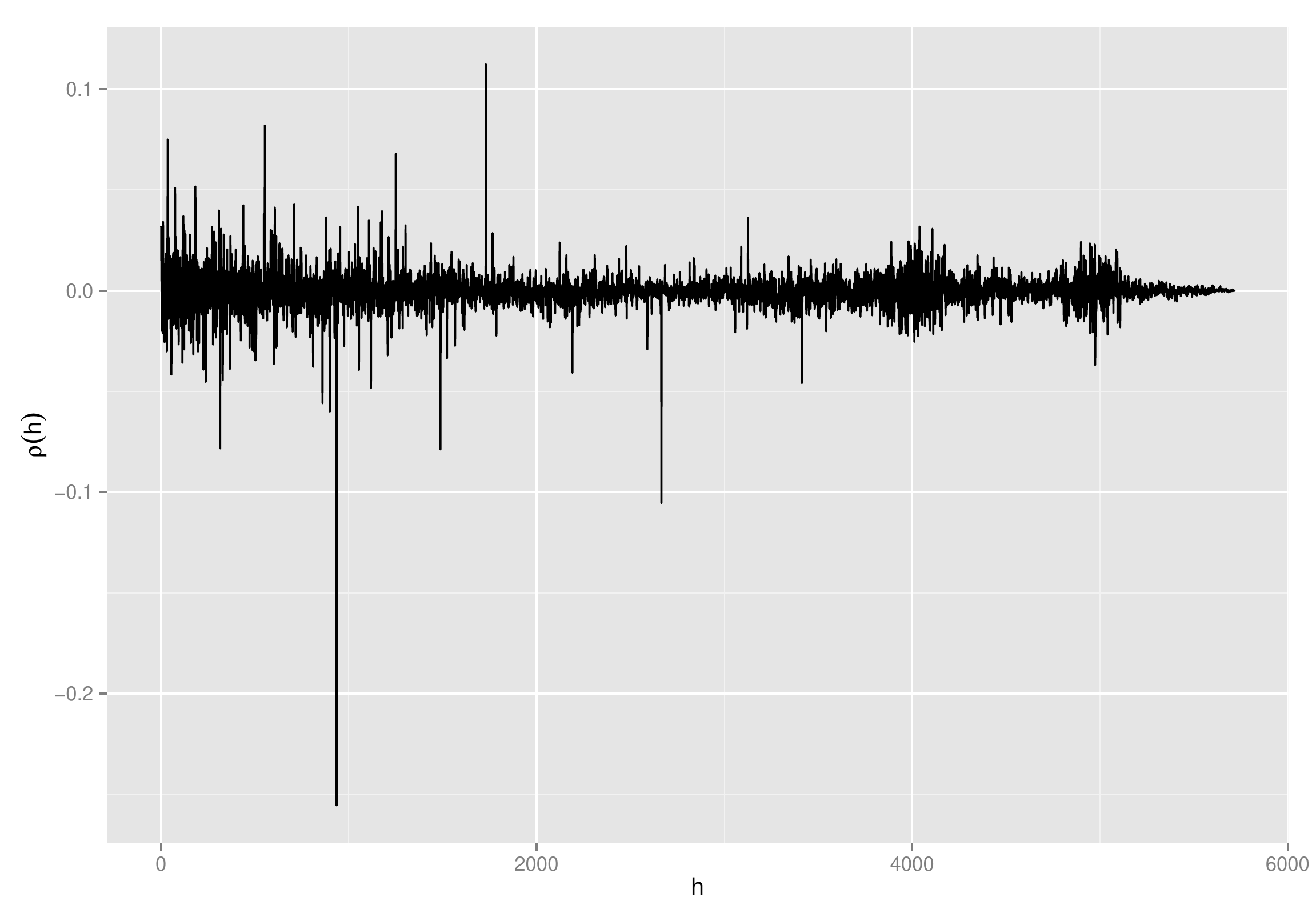}&
\includegraphics[width = 2.3in]{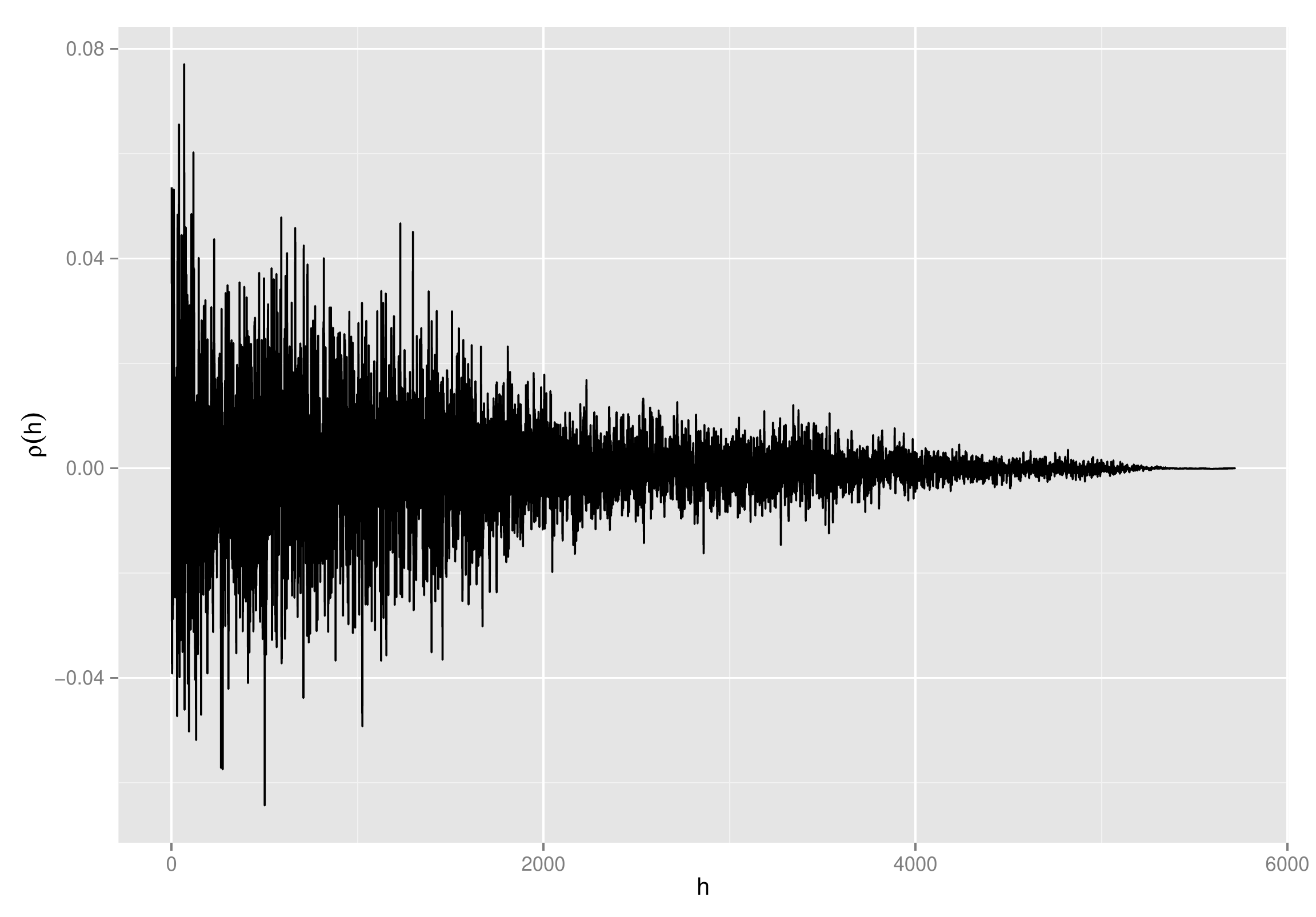}
\end{array}$
\end{center}
\caption{The autocorrelation function (\ref{autocorrelation}) as a function of the lag $h$. Left: For the forward differences between the opening and closing prices $\{L_{n}=X_{n}-Y_{n-1}\}$, mean value $\langle \rho \rangle = -8.7\times 10^{-5}$ and standar deviation $\langle\langle \rho \rangle\rangle =  0.0089$. Right: For the differences day by day $\{D_{n} = Y_{n}-X_{n}\}$, mean value $\langle \rho \rangle = -8.7\times 10^{-5}$ and standar deviation $\langle\langle \rho \rangle\rangle =  0.0095$. \label{diff-X&Y}}
\end{figure}
\end{center}

\section{Results}

The raw data used for this paper was the opening and closing prices of the MEXBOL Index in the period 
Nov/08/1991-Sept/05/2014.

The Table 1 displays the optimal values of parameter  $(\alpha, \beta, \gamma, \delta)$ for the empirical distributions $\alpha-stable$ of the returns $r$ and logarithmic returns $R$ of closing price, respectively, obtained with the function  \emph{stableFit} of the library \emph{fBasics}, implemented by \emph{GNU-R}. 

\begin{center}
\begin{table}[h]
\caption{Optimal parameters for $\alpha-$stable L\'evy distribution in returns}
\centering
\begin{tabular}{|c|c|c|c|c|}\hline
Time serie & $\alpha$&$\beta$&$\gamma$&$\delta$\\\hline
$r_{n} = Y_{n}/Y_{n-1}$ &$1.5870$&$-0.014$&$0.5148293113$&$-0.0006526977$\\\hline
$R_{n} = \ln(r_{n})$&$1.548$&$-0.041$&$0.5144696$&$0.0105618$\\\hline

\end{tabular}
\label{Table1}
\end{table}
\end{center}

\noindent With the values of Table \ref{Table1} and the equation (\ref{stable}) the asymptotic decay in power law for the normalized returns $\{r_{n}\}$ and the logarithmic returns $\{R_{n} = \ln(r_{n})\}$ are

\begin{eqnarray}\label{PowerLaw}
f_{r}(x) & = & \frac{C_{1}}{x^{1+\alpha_{r}}}\\\nonumber
f_{R}(x) & = & \frac{C_{2}}{x^{1+\alpha_{R}}}
\end{eqnarray}

\noindent with $C_{1} = 0.7354585, \alpha_{r} = 1.5870$ and $C_{2} = 0.6442726, \alpha_{R} = 1.548$. For comparison, in \cite{coronel1, coronel2, gabaix, gopikrishnan} have reported that an empirical distribution for the extreme values of the returns is a power-law type $P(\mid r \mid > x) \sim x^{-\zeta_{r}}$, with $\zeta_{r} \approx 3$.

Given a set of parameters $(\alpha, \beta, \gamma, \delta)$, using the library \emph{stabledist} \cite{wertz} the corresponding numerical values of an $\alpha-$stable distribution are obtained. The differences between the $\alpha-$stable distributions for the normalized returns and logarithmics returns of time serie of closing prices $\{Y_{n}, n = 0, 1, 2, \cdots, N\}$ are presented inf Fig. 8.

\begin{center}
\begin{figure}[htbp] 
\begin{center}$
\begin{array}{cc}
\includegraphics[width = 4.7in]{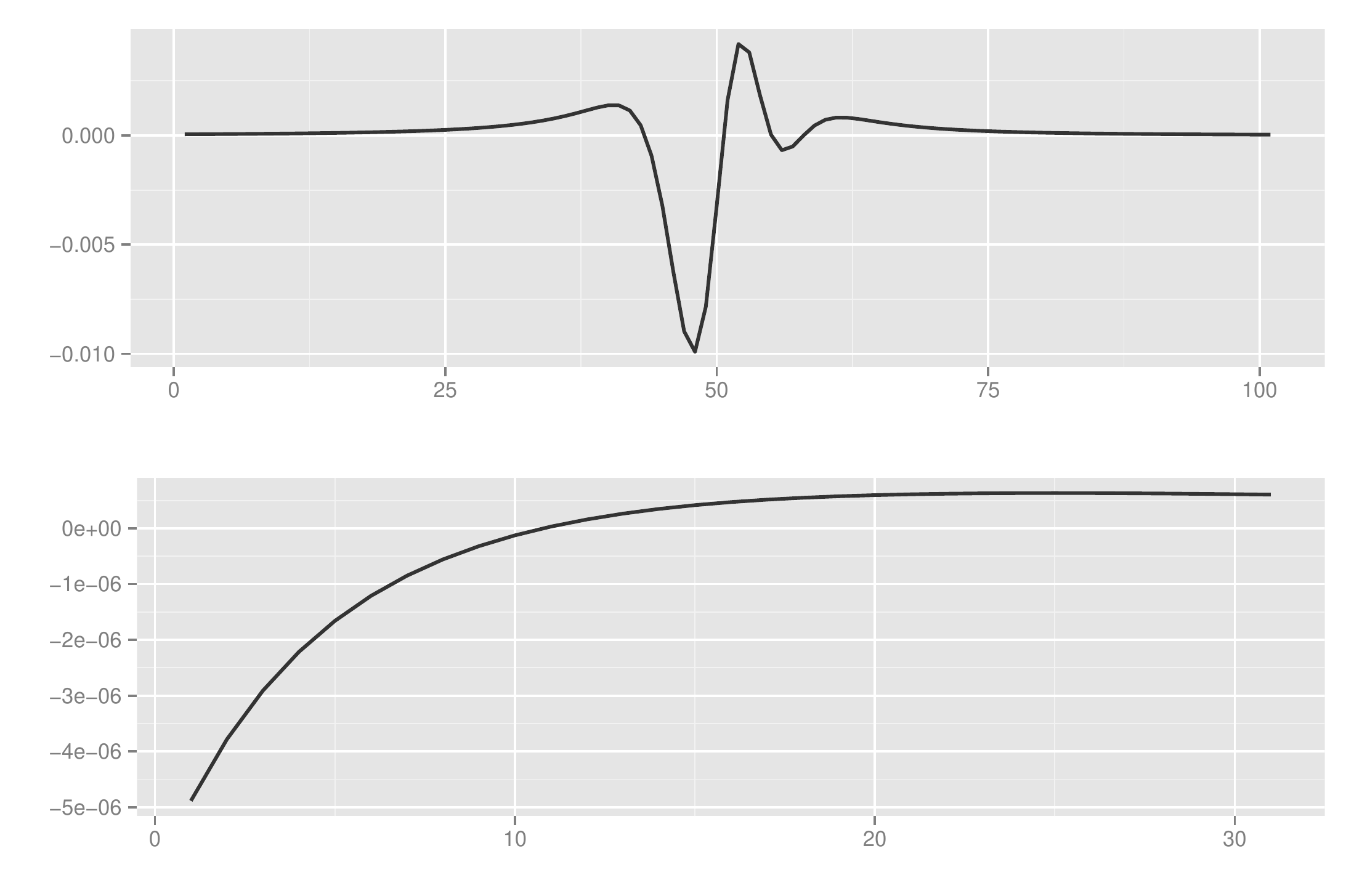}
\end{array}$
\end{center}
\caption{Top: Differences between the empirical $\alpha-$stable distributions: $\Delta = P_{R}(x)-P_{r}(x)$. Bottom: Differences for the power laws decay $\Delta = f_{R}(x)-f_{r}(x)$, equation (\ref{PowerLaw}).}
\end{figure}
\end{center}

The Table 2 shows the optimal $\alpha-$stable parameters for the time series $\{L_{n} = X_{n}-Y_{n-1}\}$ and $\{D_{n} = Y_{n}-X_{n}\}$. 

\begin{center}
\begin{table}[h]
\caption{Optimal parameters for $\alpha-$stable L\'evy distribution}
\centering
\begin{tabular}{|c|c|c|c|c|}\hline
First differences & $\alpha$&$\beta$&$\gamma$&$\delta$\\\hline
$L_{n} = X_{n}-Y_{n-1}$ &$0.757$&$0.097$&$0.05525003$&$-0.03788360$\\\hline
$D_{n} = Y_{n}-X_{n}$&$1.031$&$0.009$&$0.27469413$&$-0.03035873 $\\\hline
\end{tabular}
\label{Table2}
\end{table}
\end{center}

\noindent In the Fig. 9 at the left the empirical $\alpha-$stable density probability, corresponding to the parameters of the
Table 2, are shown. For comparison, a profile of Gaussian white noise, with $\alpha-$stable parameters
$(\alpha, \beta, \gamma, \delta) = (2, 0, 1, 0)$, is included too.

Hurst exponent for the different analized time series were estimated using different time scales, for example, daily return, return each 2 days and so on.

In the Table 3 the different values of hurst exponent calculated through $R/S$ analysis for all time series analyzed are shown.

\begin{center}
\begin{table}[h]
\caption{Hurst exponent in the IPC Index}
\centering
\begin{tabular}{|l|l|l|}
\hline\hline
$\hspace{1.3cm}$ Data	          &            & $\hspace{0.7cm} H$\\\hline
Opening prices&$X_{n}$& 1.018647\\\hline
Closing prices& $Y_{n}$   &   1.018856\\\hline
Logarithmic returns& $R_{X} = log(X_{n}/X_{n-1})$&  0.532947\\\hline
Logarithmic returns &$R_{Y} = log(Y_{n}/Y_{n-1})$ 	&    0.5318121\\\hline
Differences within& $\Delta Y_{n} = Y_{n}-Y_{n-1}$&
  0.5284645\\\hline
 Differences between &$D_{n} = Y_{n}-X_{n}$ & 0.421346\\\hline
Forward differences& $L_{n}=X_{n}-Y_{n-1}$  &0.6051881\\\hline
Gaussian white noise& $Z_{n}$&0.521465\\\hline
\end{tabular}
\label{Table3}
\end{table}
\end{center}

The Fig. 9 at the right show the asymptotic behavior of Hurst exponent when 
$n \rightarrow \infty$ for the $R/S$ analysis. When the time scale increases the autocorrelation also increases and the Hurst exponent approaches one, i.e., the effects of long-term memory are amplified.

\begin{center}
\begin{figure}[htbp]
\begin{center}$
\begin{array}{cc}
\includegraphics[scale = 0.28]{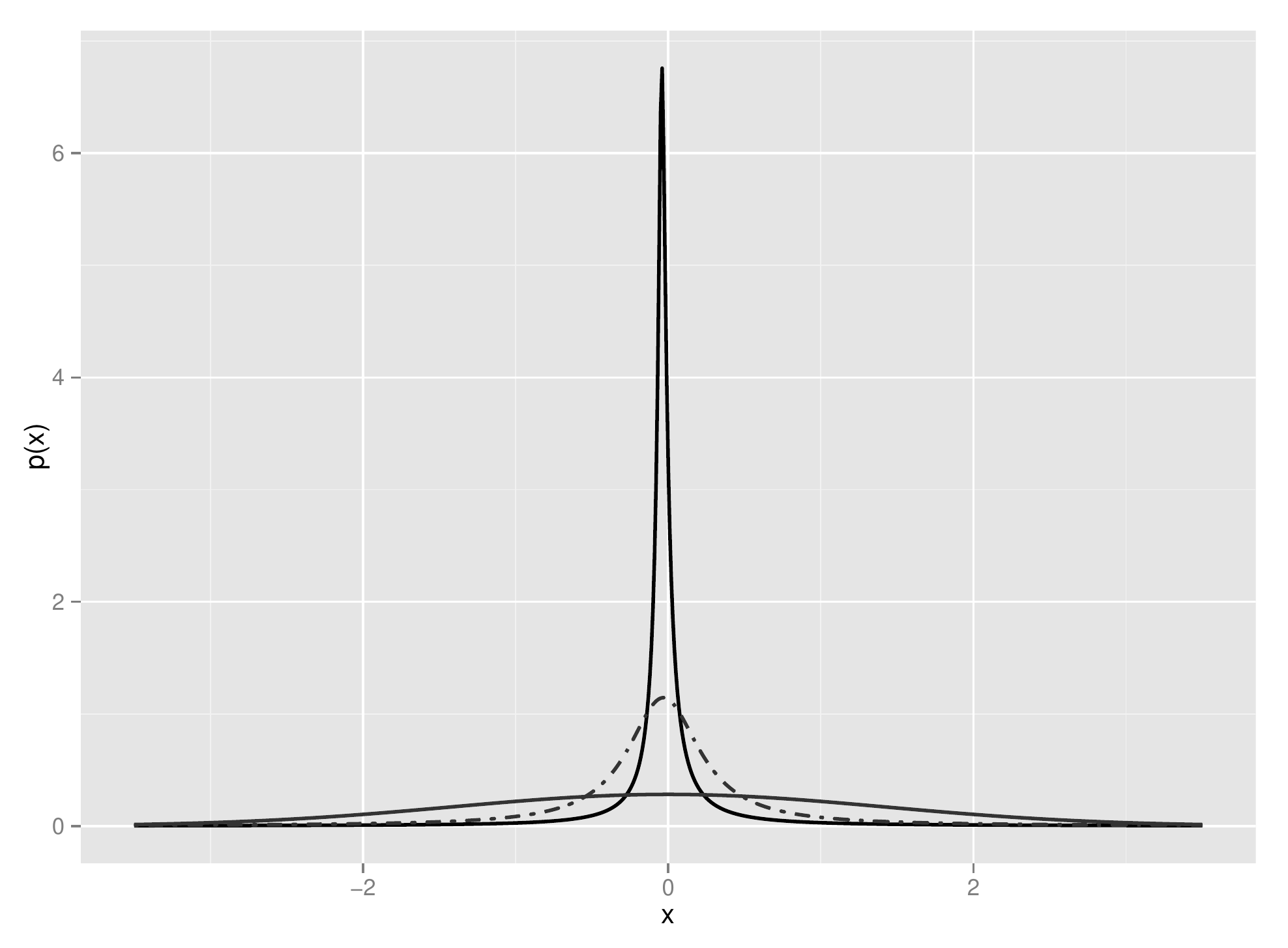}&
\includegraphics[scale = 0.25]{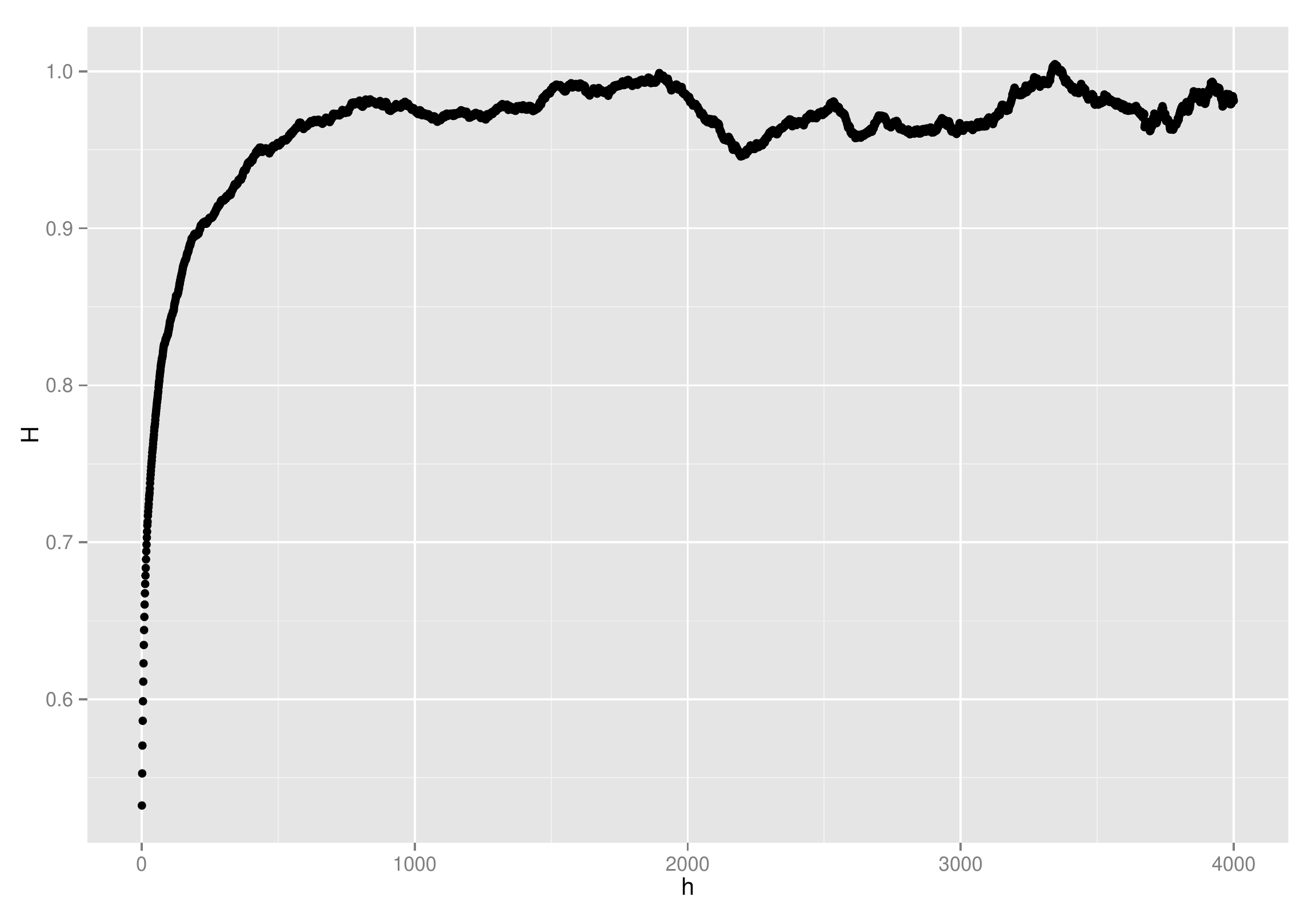}
\end{array}$
\end{center}
\caption{Left: $\alpha-$stable distribution in the forward differences $\{L_{n}=X_{n}-Y_{n-1}\}$, sharpest curve, in the differences day by day $\{D_{n} = Y_{n}-X_{n}\}$, the dashed curve, and a Gaussian white noise $\{Z_{n}\}$, most flattened curve. Rigth: Asymptotic limit in time for
Hurst exponent of the returns $\{r_{n} = Y_{n}/Y_{n-1}\}$ for the closing prices. On the horizontal axis the scale of time or lag $h$ is the size of partition in the $R/S$ analysis, ranging from $1$ to $4000$. When the scale of observation increases the correlation within the data and the trend is present. \label{hurst}}
\end{figure}
\end{center}

A summary of some observed characteristics in IPC closing prices time series are shown in Fig. 10.
In this graph 
it is possible identified at least three ``bubble
zones'' corresponding each one of them 
with a specific financial crash, 
``The effect Tequila'' of 
1994, the Asian financial crisis from 1997 to 
1998 and the recent Global financial crisis
of 2008. The big fluctuations, argues, is an indicative 
of significant correlations effects of long-term, corroborated by the form of autocorrelation function  between the opening price in the $n$-th day and the closing price in the $(n-1)$-th day, the $L_{n} = 	X_{n}-Y_{n-1}$ differences.

\begin{center}
\begin{figure}[htbp]
\begin{center}
\includegraphics[scale = 0.43, angle = 0]{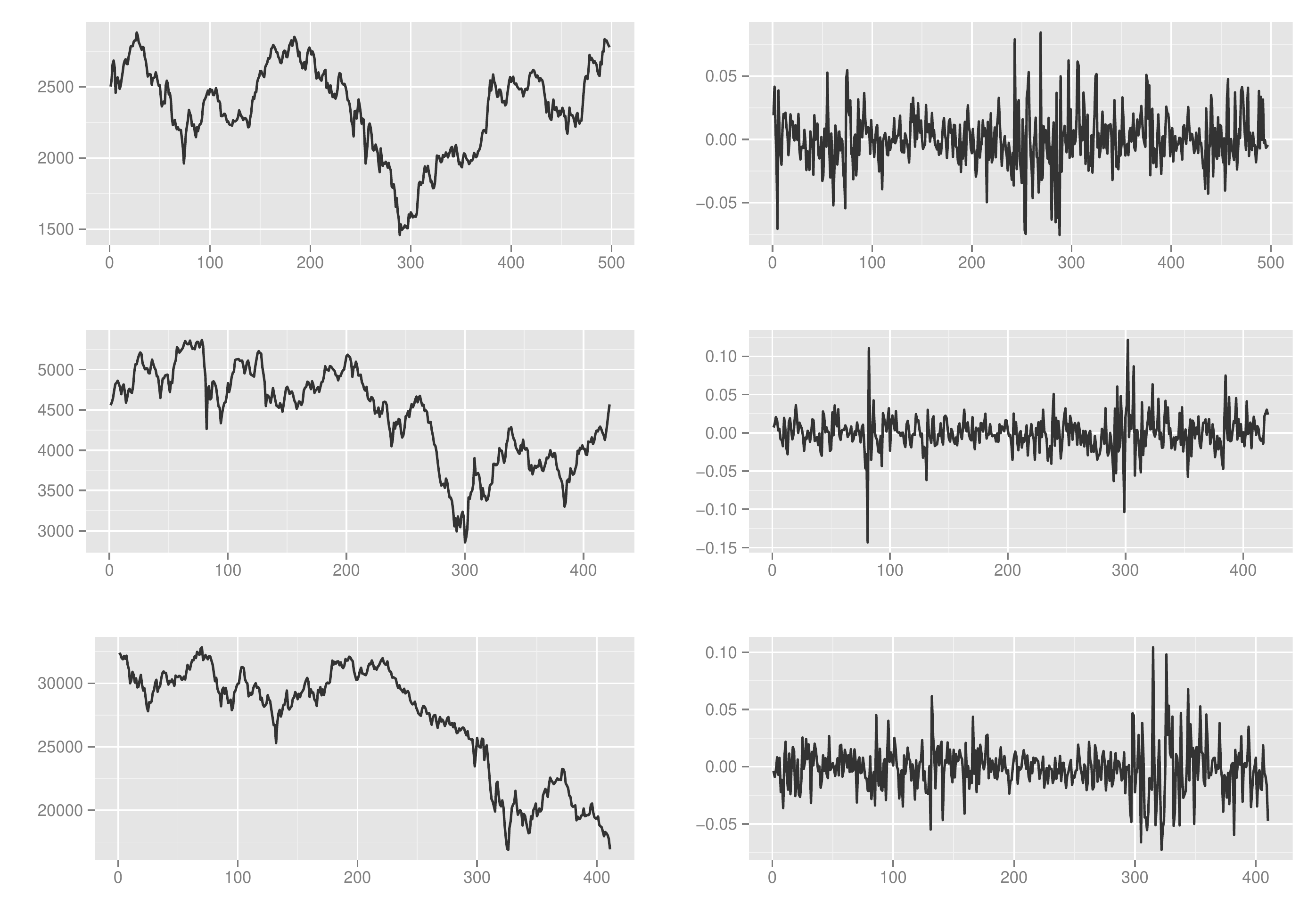}
\caption{MEXBOL. Figs. at the left: A zoomed of the three turbulent moments of closing price $\{Y_{n}\}$ 
listed in Fig. 1. Figs. at the right: The corresponding logarithmic returns $r_{n} = log(Y_{n}/Y_{n-1})$. 
Top: The effect Tequila of 
1994. In middle: Asian crisis. Bottom: Global crisis. \label{mexbol}}
\end{center}
\end{figure}
\end{center}

          \section{Conclusions}

Heavy tails in the series of the returns of opening or closing prices and in the delayed first differences between opening and closing prices, clearly indicates a non diffusive fluctuations dynamics or a non Gaussian behavior for MEXBOL IPC Index.
At the same time, this allows the presence of long term correlations and memory effects not compatible with the concept of
an efficent market. In other words, return values does not represent a simple random walk where jumps are independent with a
finite variance.

Despite returns $\{r_{n}\}$ and the logarithmic returns $\{R_{n}\}$ of closing or opening prices, even the differences $\{L_{n}\}$, exhibit a symmetric, stationary and homoscedastic behavior (Figs. 2 and 5 in top),
the series of differences $\{D_{n}\}$ (Fig. 5 in bottom) show a growing volatility over time. This sistematic increment of fluctuations size, and the average difference between closing and 
opening prices $\langle D\rangle = 7.14$  create a clear large scale positive trend in the $\{Y_{n}\}$ serie, see the Fig. 1.
As before, even though the average margin of closing prices is \emph{small} in a daily scale, with a value slightly bigger than 1,
$\langle \Delta Y /Y\rangle = 7.29/10^4$, it is \emph{big} enough to create a global growth of closing prices.

Although the Hurst exponent for the 
return of closing prices in the MEXBOL Index given by the $R/S$ are very similar and close to $1/2$ (an expected value for Gaussian white noise), the strong
difference between the time series of logarithmic returns $log(r)$ and an empirical Gaussian white noise, Figs. 2, 3, and 4, results from the presence at different
time scales of recurrent large fluctuations in returns $\{r_{n}\}$ or logarithmic returns $\log(r)$, a behavior that captures the underlying fractal nature of
many financial time series. 

The IPC Index for closing and opening prices analysis shows an incresing growth for long time period, it means that the Mexican stock market and the economic stability is a good option to investment. 

In spite of the recurrent crisis in past such as the Tequila effect, Asian crisis and the recent Global crisis, the results in this research show that the Mexican economy shown robustness  along the past 23 years.

\section{Acknowledgements}

The authors would like to thanks CONACYT, PROMEP, PAICYT and the Universidad Aut\'onoma de Nuevo Le\'on for support this research.


\end{document}